\documentclass{jfm}

\usepackage{graphicx}
\usepackage{newtxtext}
\usepackage{newtxmath}
\usepackage{natbib}
\usepackage{hyperref}
\hypersetup{
    colorlinks = true,
    urlcolor   = blue,
    citecolor  = black,
}

\newcommand{\RomanNumeralCaps}[1]
\linenumbers
\newcommand{\dd}{\:\mathrm{d}}
\newcommand{\retau}{ \Rey_{\tau} }

\newcommand{\abs}[1]{\left| #1 \right|}

\usepackage{array}
\newcolumntype{C}[1]{>{\centering\let\newline\\\arraybackslash\hspace{0pt}}m{#1}}
\usepackage{multirow,multicol}
\usepackage{hhline}
\usepackage[table]{xcolor}
\definecolor{Grey}{HTML}{A9A9A9}
\definecolor{Blue}{HTML}{005ce6}
\definecolor{Red}{HTML}{ff3300}
\definecolor{Green}{HTML}{79ff4d}

\title{Spatio-temporal characterization of non-linear forcing in turbulence}

\author{Yuting Huang\aff{1}
  \corresp{\email{yhuang1@caltech.edu}},
  Simon S. Toedtli\aff{1,2},
  Gregory P. Chini\aff{3,4}
 \and 
 Beverley J. McKeon\aff{5}}

\affiliation{
	\aff{1}Graduate Aerospace Laboratories, California Institute of Technology, Pasadena, CA 91125, USA
	\aff{2}Department of Mechanical Engineering, Johns Hopkins University, Baltimore, MD 21218, USA
    \aff{3}Program in Integrated Applied Mathematics, University of New Hampshire, Durham, NH 03824, USA
    \aff{4}Department of Mechanical Engineering, University of New Hampshire, Durham, NH 03824, USA
	\aff{5}Department of Mechanical Engineering, Stanford University, Stanford, CA 94305, USA}

\begin{document}
\maketitle

\begin{abstract}
The quadratic convection term in the incompressible Navier-Stokes equations is considered as a non-linear forcing to the linear operator, and it is studied in the Fourier domain through the analysis of interactions between triadically compatible wavenumber-frequency triplets. Interaction coefficients are proposed to quantify the contribution to the forcing by each pair of triplets and are computed using data from direct numerical simulations of a turbulent channel at $\retau \approx 550$. The coefficients show the importance of interactions involving streamwise large scales. The regions of non-linear interactions permitted under the quasi-linear (QL) and generalized quasi-linear (GQL) assumptions are shown to be significant contributors to the forcing and increasing the number of GQL-large scales is shown to monotonically increase the total forcing captured, providing a possible reason for the success of QL and GQL simulations. The tools presented are expected to be useful for improving modeling of the nonlinearity, especially in QL, GQL, and resolvent analyses, and understanding the amplitude modulation mechanism relating large-scale fluctuations to the modulation of near-wall structures.
\end{abstract}

\begin{keywords}
Turbulent boundary layers, Turbulence modelling
%Authors should not enter keywords on the manuscript, as these must be chosen by the author during the online submission process and will then be added during the typesetting process (see \href{https://www.cambridge.org/core/journals/journal-of-fluid-mechanics/information/list-of-keywords}{Keyword PDF} for the full list).  Other classifications will be added at the same time.
\end{keywords}

%{\bf MSC Codes }  {\it(Optional)} Please enter your MSC Codes here

\section{Introduction}
The beauty and complexity of turbulent fluid flow arise largely from the non-linear convective terms in the Navier-Stokes equations (NSE). In many situations, the non-linear terms remain an essential and challenging part of our understanding of turbulence. Under a classical Reynolds decomposition, i.e. the definition of turbulent fluctuations relative to a temporally- or spatio-temporally-averaged mean field, the quadratic non-linearities in the incompressible NSE manifest themselves as a convolution of triadically compatible, i.e. resonant, interactions in the Fourier domain, linking a dyad of interacting scales to nonlinearity at a third scale. Such interactions have been studied in the spatial domain~\citep[e.g.][]{Cheung14} and in the context of spatial and spectral fluxes~\citep[e.g.][]{Marati04}. The coherence between spatial scales can also be studied through the skewness of the velocity and amplitude modulation of the small scales by large ones, e.g. \cite{marusic2010}, both of which can be expressed as a measure of relative phase between modes \citep{Duvvuri_McKeon_2015}. Similarly, \citet{Schmidt_2020} proposed the bispectral mode decomposition to study coherence in the velocity signals among spatial triads and analyzed the interacting frequency components using maxima in the mode bispectrum.

One approach to reduce the complexity (and cost) of the nonlinear terms is to employ a quasi-linear (QL) model, in which the resolved nonlinear interactions are restricted to those either involving or resulting in the zero streamwise wavenumber, $k_x=0$, content \cite[e.g.][]{Farrell_Ioannou_2007}. The approach rests on the admitted interactions capturing the key elements of the nonlinearity, while the remaining unresolved interactions are neglected or approximated with a suitable model. Self-sustaining simulations can be achieved with flow features that resemble those obtained from direct numerical simulation (DNS), thus providing a cost-efficient model alternative to DNS of the full NSE. Generalized quasi-linear (GQL) analysis allows for resolving non-linear interactions involving increasing numbers of harmonics of the streamwise fundamental wavelength associated with the domain, enabling spectrally non-local streamwise energy transfers \citep{Marston_Chini_Tobias_2016}; the complexity of the simulations increases as the filter separating large and small scales moves to smaller scale, converging to DNS in the limit of no filter. To our knowledge, the importance of the resolved nonlinear interactions relative to the unresolved (neglected or modeled) ones in QL and GQL has not been fully characterized or explained, which is one motivation for the present work.

Other modeling approaches include linear analyses such as resolvent analysis \citep{mckeon2010}, which is used to identify spatio-temporal basis sets for the nonlinearity, treated as an input forcing to the linear NSE (resolvent) operator, that is preferentially amplified by the system and the associated system response. The nonlinearity may be treated crudely as a broadband input to analyze the properties of the linear operator or with more sophistication as a colored forcing arising from nonlinear modal interactions to address the turbulence closure problem~\citep{mckeon2017}. Some characteristics of the forcing in resolvent representations have been analyzed by, e.g.,
\citet{Morra_Nogueira_Cavalieri_Henningson_2021}. \cite{Karban_NL_23} have investigated the key triads underpinning minimal Couette flow. Understanding the contribution of individual triads to overall forcing within the resolvent framework as well as within the spatio-temporal model of the energy cascade is the second objective of this work.

\section{Formulation}\label{sec:formulation}
We consider incompressible, turbulent channel flow and use a Reynolds decomposition of state variables $\boldsymbol{\hat{q}}(x,y,z,t)$ to define a spatio-temporal mean profile $Q(y)$, averaged in $x$, $z$, and $t$, and the perturbations $\boldsymbol{q}(x,y,z,t)$ around the mean. A Fourier decomposition is then employed in the homogeneous directions of $x$, $z$, and $t$:
\begin{equation}
	\boldsymbol{q}(\boldsymbol{x},t) = \iiint_{-\infty}^{\infty} \boldsymbol{q}(\boldsymbol{k},y) e^{i(k_x x+ k_z z - \omega t)} \dd k_x \dd k_z \dd \omega ,
\end{equation}
where we have introduced the spatial vector $\boldsymbol{x} = [x,y,z]$, and a wavenumber-frequency triplet $\boldsymbol{k} = [k_x, k_z, \omega]$. Here $k_x$, $k_z$ are the streamwise, and spanwise wavenumbers and $\omega$ is the temporal frequency. %Therefore, $q(\boldsymbol{k},y)$ represents a state variable after Fourier transform in all 3 homogeneous directions of $x$, $z$, and $t$.

\subsection{Non-linear forcing in the Navier-Stokes equations}
The non-linear (quadratic) terms $\boldsymbol{f}$ are defined in physical space by:
\begin{equation}
	\boldsymbol{f}(\boldsymbol{x},t) =  - \boldsymbol{u}(\boldsymbol{x},t) \cdot \nabla \boldsymbol{u}(\boldsymbol{x},t),
\end{equation}
while in (discrete) Fourier space the non-linear forcing at a wavenumber-frequency triplet $\boldsymbol{k}_3$ can be written in terms of a convolution of the velocity fields at $\boldsymbol{k}_1$ and $\boldsymbol{k}_2$:
\begin{equation}
	\boldsymbol{f}(\boldsymbol{k}_3,y) = - \sum_{ \boldsymbol{k}_{1}+\boldsymbol{k}_{2} = \boldsymbol{k}_3}  \boldsymbol{u}(\boldsymbol{k}_1,y) \cdot \nabla  \boldsymbol{u}(\boldsymbol{k}_2,y). \label{eq:fconv}
\end{equation}
The requirement of $\boldsymbol{k}_1 + \boldsymbol{k}_2 = \boldsymbol{k}_3$ is the triadic compatibility or resonance constraint. Here $\boldsymbol{u}$ is the fluctuation velocity vector.

In the resolvent formulation, the Fourier-transformed NSE are written in an input-output form, where $\boldsymbol{f}(\boldsymbol{k},y)$ is considered as an input forcing to the resolvent operator $\mathcal{H}(\boldsymbol{k},y)$:
\begin{equation}
	\boldsymbol{q}(\boldsymbol{k},y)=\begin{bmatrix} \boldsymbol{u}(\boldsymbol{k},y) \\ p(\boldsymbol{k},y) \end{bmatrix} = \mathcal{H}(\boldsymbol{k},y) ~ \boldsymbol{f}(\boldsymbol{k},y), \label{eq:resolvent_nse}
\end{equation}
where $p$ is the pressure fluctuation. The forcing is responsible for the distribution of energy between different scales.

\subsection{Interaction Coefficients}
The contribution from the interaction between a pair of triplets $\boldsymbol{k}_1$ and $\boldsymbol{k}_2$ to the resulting forcing at $\boldsymbol{k}_3 = \boldsymbol{k}_1 + \boldsymbol{k}_2$, can be quantified through an interaction coefficient $P(\boldsymbol{k}_{1}, \boldsymbol{k}_{2})$:
\begin{equation}
	P(\boldsymbol{k}_{1}, \boldsymbol{k}_{2}) =  \int -\boldsymbol{u}(\boldsymbol{k}_1,y) \cdot \nabla  \boldsymbol{u}(\boldsymbol{k}_2,y)  \cdot \boldsymbol{f}^*(\boldsymbol{k}_1 + \boldsymbol{k}_2,y)  \dd y, \label{eq:defp}
\end{equation}
which is the inner product between $-\boldsymbol{u}(\boldsymbol{k}_1,y) \cdot \nabla  \boldsymbol{u}(\boldsymbol{k}_2,y)$ and $\boldsymbol{f}(\boldsymbol{k}_3,y)=\boldsymbol{f}(\boldsymbol{k}_1 + \boldsymbol{k}_2,y)$. The intentionally un-normalized coefficients take the forcing magnitude into consideration, e.g. a large contribution to a small magnitude forcing is treated as unimportant. Note that these coefficients differ from those, e.g. in \cite{mckeon2017}, defined for the interactions between resolvent modes instead of between DNS Fourier modes. To facilitate computation and visualization of $P(\boldsymbol{k}_{1}, \boldsymbol{k}_{2})$, we define $P_{k_x}$ and $P_{\omega}$ by summation in 4 of the 6 dimensions:

\begin{subeqnarray}
	P_{k_x}(k_{x1}, k_{x2}) &=&  \sum_{k_{z1}}\sum_{k_{z2}}\sum_{\omega_1}\sum_{\omega_2}  P(\boldsymbol{k}_{1}, \boldsymbol{k}_{2}), \label{eq:defpkx}\\
%	P_{k_z}(k_{z1}, k_{z2}) &=&  \sum_{k_{x1}}\sum_{k_{x2}}\sum_{\omega_1}\sum_{\omega_2}  \int -\boldsymbol{u}(\boldsymbol{k}_1,y) \cdot \nabla  \boldsymbol{u}(\boldsymbol{k}_2,y)  \cdot \boldsymbol{f}^*(\boldsymbol{k}_1+\boldsymbol{k}_2,y) \dd y, \label{eq:defpkz}\\
	P_{\omega}(\omega_1, \omega_2) &=&  \sum_{k_{x1}}\sum_{k_{x2}}\sum_{k_{z1}}\sum_{k_{z2}}  P(\boldsymbol{k}_{1}, \boldsymbol{k}_{2}). \label{eq:defpom}
\end{subeqnarray}
$P_{k_x}$ defined in equation~\eqref{eq:defpkx}(\textit{a}) describes the interaction in the streamwise direction between $k_{x1}$ and $k_{x2}$, summed over all possible $k_z$ and $\omega$ interactions. Similarly, $P_{\omega}$ describes the interaction between $\omega_1$ and $\omega_2$. $P_{k_z}(k_{z1}, k_{z2})$ can be similarly defined but is not explored here.

Equivalent forms for the two interaction coefficients can be obtained in the $(z,t)$ and $(x,z)$ domains with a significant reduction in computation costs and memory requirements:
\begin{eqnarray}
	P_{k_x}(k_{x1}, k_{x2}) &=&  - \frac{1}{{N_z N_t}} \sum_{z} \sum_{t}  \int \boldsymbol{u}(k_{x1},y,z,t) \cdot \nabla\boldsymbol{u}(k_{x2},y,z,t)\cdot \boldsymbol{f}^*(k_{x1}+k_{x2},y,z,t) \dd y, \nonumber\\
%	P_{k_z}(k_{z1}, k_{z2}) &=&  - \frac{1}{{N_x N_t}} \sum_{x} \sum_{t}  \int \boldsymbol{u}(x,y,k_{z1},t) \cdot \nabla\boldsymbol{u}(x,y,k_{z2},t)\cdot \boldsymbol{f}^*(x,y,k_{z1}+k_{z2},t) \dd y, \nonumber\\
	P_{\omega}(\omega_{1}, \omega_{2}) &=&  - \frac{1}{{N_x N_z}} \sum_{x} \sum_{z}  \int \boldsymbol{u}(x,y,z,\omega_1) \cdot \nabla\boldsymbol{u}(x,y,z,\omega_2)\cdot \boldsymbol{f}^*(x,y,z,\omega_1+\omega_2) \dd y, \nonumber \\ \label{eq:pkx_alt}
\end{eqnarray}
Note that $\boldsymbol{u}(k_{x1},y,z,t)$ here denotes the velocity Fourier transformed only in $x$, and differs from $\boldsymbol{u}(\boldsymbol{k}_1,y)$ in equation~\eqref{eq:defp}, which is Fourier transformed in $x$, $z$ and $t$. These forms provide alternative interpretations of the projection coefficients: $P_{k_x}(k_{x1}, k_{x2})$ can now be understood as quantifying the importance of the interaction between $k_{x1}$ and $k_{x2}$ averaged in $z$ and $t$ as well as $y$. Similarly, $P_{\omega}$ reflects an average in $x$, $y$ and $z$. $P_{k_x}$ can be computed in the temporal domain, while $P_{\omega}$ requires a Fourier transform in time, the implementation of which is discussed further below.

Using the Hermitian symmetry of the velocity Fourier coefficients, it can be shown that the interaction coefficients are also Hermitian functions. However, they are asymmetric about their two arguments, due to the action of the velocity gradient tensor. For example, $P_{k_x}$ satisfies:
\begin{equation}
	P_{k_x}(k_{x1}, k_{x2}) = P_{k_x}^*(-k_{x1}, -k_{x2})\neq P_{k_x}(k_{x2}, k_{x1}). \label{eq:hermitiansymmetry}
\end{equation}
We retain the two separate coefficients associated with each combination of $(k_{x1}, k_{x2})$ to maximize the information about the dominant interactions within each triad, which would be lost under a combined coefficient. 
%However, the symmetric part of $P_{k_x}$ for example, denoted as $\hat{P}_{k_x}$, can be simply computed by
%\begin{equation}
%	\hat{P}_{k_x}(k_{x1}, k_{x2}) = \frac{1}{2} \bracket{P_{k_x}(k_{x1}, k_{x2}) + P_{k_x}(k_{x2}, k_{x1})}.
%\end{equation}

\subsection{Relation to triple correlation and bispectrum}
The interaction coefficients studied here may be related to the more usual triple correlation and bispectrum. Following \citet{Lii_Rosenblatt_Atta_1976}, the three-point spatial triple correlation for three state variables $q_l$, $q_m$ and $q_n$, can be defined as:
\begin{equation}
	R_{lmn}(\boldsymbol{r},\boldsymbol{r'}) = \langle q_l(\boldsymbol{x})q_m(\boldsymbol{x}+\boldsymbol{r})q_n(\boldsymbol{x}+\boldsymbol{r'}) \rangle_{\boldsymbol{x}},
\end{equation}
where $\langle \cdot \rangle_{\boldsymbol{x}}$ represents a spatial average. Two triple spatial Fourier transforms of $R_{lmn}(\boldsymbol{r},\boldsymbol{r'})$ in $\boldsymbol{r}, \boldsymbol{r'}$ lead to the three-dimensional spatial bispectrum:
\begin{equation}
	B_{lmn}(\boldsymbol{\hat{k}}, \boldsymbol{\hat{k}'}) = q_l^*(\boldsymbol{\hat{k}}+\boldsymbol{\hat{k}'})q_m(\boldsymbol{\hat{k}})q_n(\boldsymbol{\hat{k}'}).
\end{equation}
The interaction coefficients proposed in this work in equation~\eqref{eq:defp} can be seen as a spatio-temporal extension of the bispectrum, 
considering the spatio-temporal wavenumbers, $\boldsymbol{k}$, that are suitable for the analysis of wall-bounded flows with homogeneous directions of $x$, $z$ and $t$ rather than simply spatial wavenumbers, $\boldsymbol{\hat{k}}$. An additional difference is that interaction coefficients in this work are specifically designed to study the contribution of an interacting pair on the resulting forcing, which to our knowledge has not been studied before. The three contributing terms to the interaction coefficients are the velocity $\boldsymbol{u}$, velocity gradient $\nabla\boldsymbol{u}$, and forcing $\boldsymbol{f}$ in contrast to previous studies that focused more on the three terms being the same flow quantity, for example, $\p u/\p x$ in \citet{Lii_Rosenblatt_Atta_1976} and velocity $\boldsymbol{u}$ in \citet{Schmidt_2020}. 

\subsection{Data from Direct Numerical Simulation} \label{sec:dns}
We evaluate the interaction coefficients described above using data from the DNS code of \citet{Flores_Jimenez_2006}, which uses a spectral discretization in the streamwise ($x$) and spanwise ($z$) directions. The channel half height is denoted as $h$, the domain size is $4\pi h \times 2 \pi h$ and the friction Reynolds number $\retau = u_{\tau} h/\nu$ is approximately 552. Turbulence statistics are in good agreement with previous studies (not shown). Quantities normalized with inner-units, using the viscous length scale $\delta_{\nu} = \nu / u_{\tau}$ and friction velocity $u_\tau$ are indicated with superscripts `$+$'. Otherwise, normalization is performed with channel center-line velocity and channel half-height. The maximum wavenumbers retained by the DNS are $k_x = \pm 127.5$, $k_z = \pm 255$, and the time sampling interval is on average 0.102 with a standard deviation of 0.003 due to the non-constant time step size. DNS snapshots are then linearly interpolated onto a uniform time vector with a sampling interval of $\Delta t = 0.1227$ for a total of 1024 uniformly spaced snapshots. The computation of $P_{\omega}$ requires a Fourier transform in time with the phase information preserved, precluding the use of Welch's method with faster convergence. FFTs in time are performed on the linearly interpolated snapshots with a Hamming window applied, resulting in a maximum temporal frequency of $\omega = \pm 25.6$ with a frequency resolution of $0.05$; very little variation was observed when  a range of different window functions were applied. To ensure the proper convergence of $P_{\omega}$, the computation is repeated using different numbers of snapshots ranging from 512 to the full 1024. Due to the difference in frequency resolution, the high-resolution results are down-sampled for comparison. 512 snapshots produce results visually indistinguishable from the full computation, while values at individual pairs of $(\omega_{1}, \omega_{2})$ take longer to converge. Very weak signatures of temporal aliasing are observed in the power spectra of the forcing at high $k_x$ but are expected to have very little effect on $P_{\omega}$.

\begin{figure}
	\centering
	\includegraphics[width=384pt]{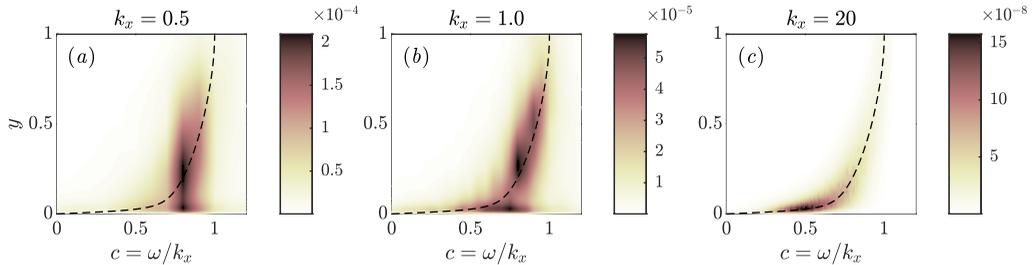}
	\caption{Contour plots of streamwise power spectra as a function of $k_x$, $y$, and the wavespeed $c = \omega/k_x$. Figures (\textit{a})-(\textit{b}) are the large scales with $k_x = 0.5, 1$, and (\textit{c}) is the small scale with $k_x = 20$. The black dash lines are the 1D mean streamwise velocity profile and are also the critical layer locations.}
	\label{fig:spec} 
\end{figure}

In figure~\ref{fig:spec}, streamwise power spectra $\phi_{uu}(c,y;k_x)$ are plotted as a function of $k_x$, $y$, and the wavespeed $c = \omega/k_x$ for two large scales: $k_x = 0.5, 1$ in figure~\ref{fig:spec}(\textit{a}) and (\textit{b}), and a representative small scale at $k_x = 20$ in figure~\ref{fig:spec}(\textit{c}). The black dashed lines in each subplot show the streamwise mean velocity profile, $U(y)$.
% DON'T THINK WE NEED THIS HERE and are also showing the critical layer, which is defined as the $y$ locations where the local streamwise mean velocity matches the wavespeed $c$. 
The energy contained in the large scales is distributed in $y$, extending almost throughout the entire channel height, and predominantly located at high wavespeeds. On the other hand, the small scales (large $k_x$) have most of the energy located at smaller wavespeeds and have an energy distribution that is mostly localized near the wall and in a $y$ range that is centered around the local mean velocity. 

\section{Spatio-temporal characteristics of the nonlinear interactions}\label{sec:results}
\begin{figure}
	\centering
	\includegraphics[width=383pt]{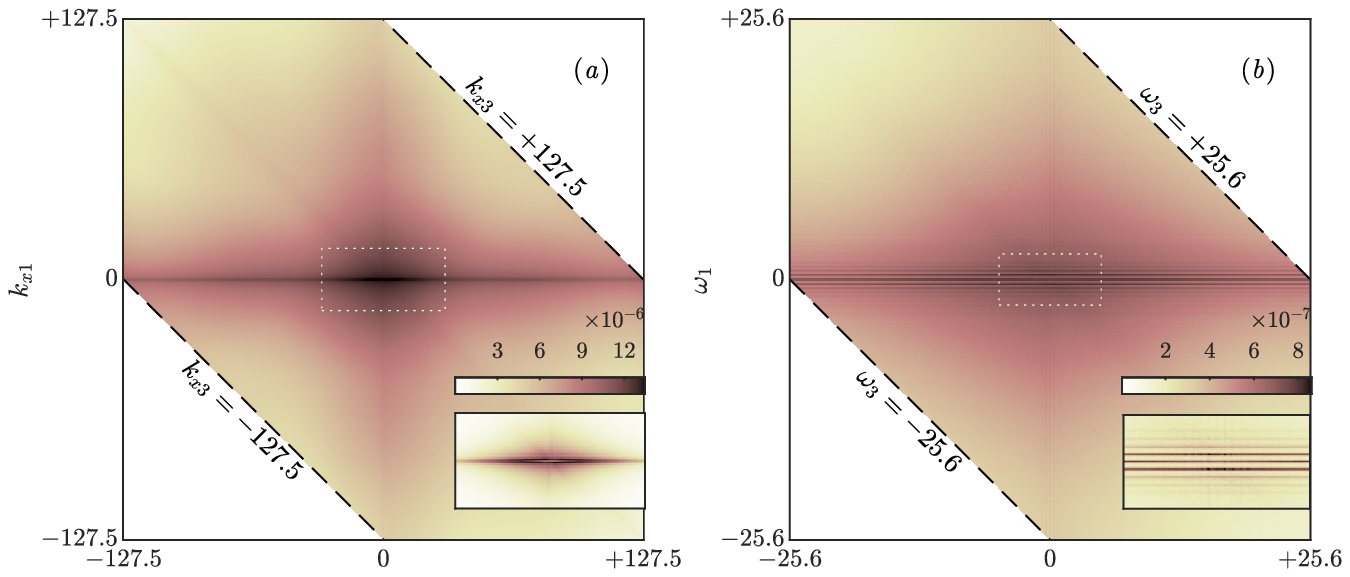}\\
	\vspace*{0.4cm}
	\includegraphics[width=383pt]{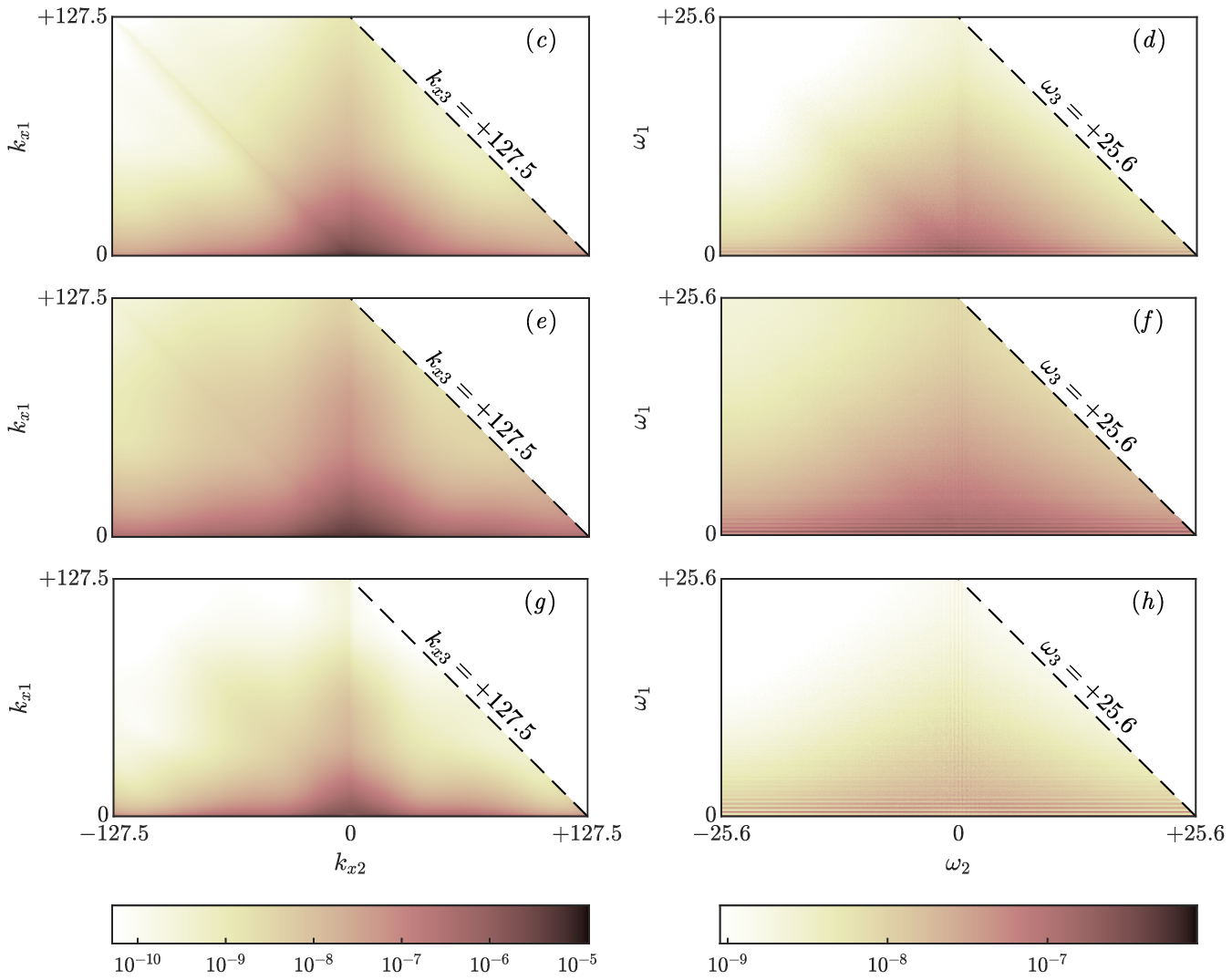}
	\caption{Heatmaps of the magnitude of the streamwise and temporal interaction coefficients with logarithmically-scaled colorbars. The inserts in \textit{(a,b)} correspond to a linear scale representation in the dashed white boxes. Left column: $\abs{P_{k_x}(k_{x1}, k_{x2})}$; right column: $\abs{P_{\omega}(\omega_{1}, \omega_{2})}$. The $y$-integration limits in equation~\eqref{eq:pkx_alt} are: \textit{(a, b)} all $y^+$; \textit{(c, d)} $y^+ \in (0, 30)$; \textit{(e,f)} $y^+ \in (30, 200)$; and \textit{(g, h)} $y^+ \in (200, 550)$. The streamwise wavenumber and temporal frequency for the velocity fields, $k_{x1}, \omega_1$, are on the vertical axis; $k_{x2}, \omega_2$ for the velocity gradient are on the horizontal axis; $k_{x3} = k_{x1} + k_{x2}$ and $\omega_3 = \omega_1 + \omega_2$ for the resulting forcing, are constant along lines with a slope of $-1$. The colorbars at the bottom are for each column of figures respectively.} %For the $y$ variation plots, only the top halves of the figures with $k_{x1},\omega_1 \geq 0$ are shown due to the Hermitian symmetry given in equation~\eqref{eq:hermitiansymmetry}.}
	\label{fig:pkxom}
\end{figure}

We consider the magnitude of the streamwise interaction coefficient $\abs{P_{k_x}(k_{x1}, k_{x2})}$ and the temporal frequency interaction coefficient $\abs{P_{\omega}(\omega_{1}, \omega_{2})}$ in the left and right columns of figure~\ref{fig:pkxom}, respectively. The phases of the coefficients, not considered here, would provide information about the type of contribution, i.e. constructive or destructive interference. Different rows correspond to different $y$ integration ranges: all $y$, followed by limits corresponding loosely to the near-wall, overlap, and wake regions. 

The streamwise wavenumber and temporal frequency for the velocity fields, $k_{x1}, \omega_1$, are on the vertical axis, and $k_{x2}, \omega_2$ for the velocity gradient are on the horizontal axis. Diagonal lines with a slope of $-1$ correspond to constant $k_{x3} = k_{x1} + k_{x2}$ and $\omega_3 = \omega_1 + \omega_2$ for the resulting forcing; the extremum values of $k_{x3} = \pm 127.5$ and $\omega_3 = \pm 25.6$ reflect the maximum $k_x$ and $\omega$ retained by the DNS. Four quadrants of $(k_{x1}, k_{x2})$ and $(\omega_{1}, \omega_2)$ are shown in  figure~\ref{fig:pkxom}(\textit{a,b}) for completeness, while the symmetry discussed above is exploited for subsequent panels in which only $k_{x1},\omega_1 \geq 0$ is shown. To highlight the details of $P$, the same logarithmic color scale is used throughout the figure. It should be noted that part of the differences in magnitude between the rows is simply attributed to the different sizes of the integration domains.

\subsection{Triad interactions reveal the importance of the large scales}
%\subsection{Interactions involving large scales}

A linear scale for the magnitude of $P$ (insets in figure~\ref{fig:pkxom}(\textit{a,b})) reveals the dominant contributions of horizontal bands involving low $k_{x1}$ and $\omega_1$. In the case of $\abs{P_{k_x}(k_{x1}, k_{x2})}$, this corresponds to the interaction between the streamwise large-scale velocity modes with velocity gradients at all scales. The high values in this region are not surprising, as the streamwise large-scale structures at $k_x = 0$ and $0.5$ are the most energetic modes in the flow field. In addition, these modes have a tall wall-normal extent as shown in figure~\ref{fig:spec}, and therefore are capable of interacting with modes of any size, centered at any $y$ location. Furthermore, the values in this horizontal region decay relatively slowly as $\abs{k_{x2}}$ increases. This is due to the compensation by the gradients, despite the velocity modes becoming less energetic with increasing $\abs{k_{x2}}$. For large $\abs{k_{x2}}$, i.e. the left and right parts of this horizontal region, the nonlinear interactions are between large-scale velocity modes and small-scale velocity gradients, generating forcing at small scales (large $\abs{k_{x3}}$). Relatively high values for these long-range interactions (in $k_x$) indicate coherence between the large and small scales, consistent with the superposition and modulation mechanisms observed in \citet{marusic2010}. For simulations with larger domain lengths in $x$, it is likely that $k_x$ values between 0 and 0.5 will replace $k_x = 0$ as the most energetic wavenumber, slightly altering the peak location of the horizontal band without significantly changing its overall structure.

Significantly lower values of $\abs{P_{k_x}(k_{x1}, k_{x2})}$are observed around the vertical axis, $k_{x2}=0$, where the modes contributing to the velocity and velocity gradient are reversed in equation~\eqref{eq:defpkx}. Spatial derivatives for the large scale are weaker and the energy of the small scales is lower than for the other case, underscoring the asymmetry of interactions within a given triad.

Finally,  a weak signature of interactions between two similar size modes generating a streamwise large-scale forcing can be observed for $k_{x3} \approx 0$. Within this diagonal region, interactions between two large scales are significantly stronger than in the corner regions, which correspond to interactions between two small scales. Consistent with the findings of \citet{Morra_Nogueira_Cavalieri_Henningson_2021}, large-scale forcing modes are mostly the result of interactions by large-scale structures, although with a weak influence from smaller modes. These smaller modes can be identified as being associated with the near-wall region by varying the $y$-integration limits, as shown in figures~\ref{fig:pkxom}(\textit{c, e, g}): the most significant differences are observed for small scales (in the top-left corner). Interactions between two small scales only contribute to the large-scale forcing near the wall but play almost no role in the outer regions.

%\bjm{PLEASE CHECK THIS PARAGRAPH MAKES SENSE} Outside of the regions described thus far, the magnitude of $\abs{P_{k_x}(k_{x1}, k_{x2})}$ falls off rapidly, dropping by orders of magnitude in the near-wall and wake regions.  Interestingly, the reduction in magnitude is smaller in the overlap region, figure~\ref{fig:pkxom}(\textit{c}), reflecting broadband but low-level contributions to the forcing.

The most prominent features observed in the frequency interaction coefficients plotted in the right column of figure~\ref{fig:pkxom} are the discrete high-value lines located horizontally around $\omega_1\approx0$. The discreteness in $\omega$ is due to the discreteness in the streamwise wavenumber $k_x$, an artifact of the finite simulation domain length, as similarly demonstrated in \citet{Gomez2014}. The frequency can be related to the phase speed for a given $\boldsymbol{k}$ via $\omega = c\cdot k_x$; at a given wavespeed, $c$, an increase in $k_x$ to the next discrete wavenumber, with an increment of 0.5 fixed by the domain length, will cause $\omega$ to increase by $0.5c$. Therefore, increments in $\omega$ are largest for large scales with high wavespeeds, appearing as discrete bands, and smaller for small scales with low wavespeeds, reflected in the smoother variation in this region. 

A more detailed analysis reveals that the 5 most prominent horizontal bands observed in figure~\ref{fig:pkxom}(\textit{b}) are located at $\omega=0$, $\pm 0.4$, and $\pm0.8$. These bands correspond well with the energetic modes at $k_x=0$, $k_x=\pm 0.5$ (with wavespeeds of $c = \omega/k_x \approx0.8$ shown in figure~\ref{fig:spec}(\textit{a})) and $k_x= \pm 1$ (with $c \approx0.8$ shown in figure~\ref{fig:spec}(\textit{b})) respectively. Furthermore, by observing the relative intensity between the smooth background and discrete lines across different $y$ locations, it can be seen that the former is more prominent near the wall in figure~\ref{fig:pkxom}(\textit{d}), while away from the wall in figure~\ref{fig:pkxom}(\textit{h}) the opposite is true. Combined with the tall $y$ extent for large scales and the concentration of energy near the wall for small scales, as demonstrated in figure~\ref{fig:spec}, these confirm that the smooth background mostly shows the triadic interactions between the small scales, while the discrete lines are mostly the result of interactions involving the large scales. Although the discreteness is an artifact of the finite simulation domain length, interactions involving the large scales are expected to be important regardless of the domain length.

% The spanwise interaction coefficient $P_{k_z}(k_{z1}, k_{z2})$, which is not shown here behaves similarly to the streamwise coefficient, with the exception that the horizontal region no longer shows a single band located around $k_{x1}\approx0$, but instead shows a dual-band structure located at $k_{z1} \approx\pm3$. This is as expected since the large-scale structures, which are the most energetic modes in the flow field, with $k_x = 0$ and $k_x = 0.5$ are also most energetic at $k_z \approx \pm 3$.

%\begin{figure}
%	\centering
%	\includegraphics[width=371pt]{{D:/Academic/Caltech/Research/0Writting/2022-08ForcingProjection/Figures/TI1}.eps}
%	\caption{Feynman diagrams depicting the important triadic interactions observed in the streamwise interaction coefficient. Subplots (\textit{a}), (\textit{b}), (\textit{c}) are for the horizontal, vertical, and diagonal regions of high magnitudes shown in figure~\ref{fig:pkx1}(b) respectively.}
%	\label{fig:TI1}
%\end{figure}

\subsection{Quasi-linear and generalized quasi-linear contributions to the forcing}
In figure~\ref{fig:pkxom}(\textit{a}), we observed three regions of dominant contributions to the forcing, all corresponding to triadic interactions involving the streamwise large scales, consistent with the assumptions underlying QL and GQL analysis. To further investigate the implications of this observation, we decompose the velocity $\boldsymbol{u}$ and non-linear forcing $\boldsymbol{f}$ in a manner reflecting the QL and GQL restrictions~\citep[e.g.][]{Marston_Chini_Tobias_2016}:
\newcommand*\centermathcell[1]{\omit\hfil$\displaystyle#1$\hfil\ignorespaces}
\begin{alignat}{7}
	\boldsymbol{u} &=&& \centermathcell{\boldsymbol{\bar{u}}}&&+&&\centermathcell{\boldsymbol{\tilde{u}}}&&+&&\centermathcell{\boldsymbol{u'}}\\
	\boldsymbol{f} &=&& \underbrace{\boldsymbol{\bar{f}}}_{k_x = 0}&&+&&\underbrace{\boldsymbol{\tilde{f}}}_{0<\abs{k_x}\leq \Lambda} &&+&& \underbrace{\boldsymbol{f'}}_{\abs{k_x}>\Lambda}.
\end{alignat}
Here $\boldsymbol{\bar{u}}, \boldsymbol{\bar{f}}$ are the streamwise averages, i.e. all modes with $k_x=0$. 
$\boldsymbol{\tilde{u}}$ and $\boldsymbol{\tilde{f}}$ contain the large scales with $k_x$ less than or equal to the cut-off $\Lambda$, i.e. $0<\abs{k_x}\leq \Lambda$, and $\boldsymbol{u'},\boldsymbol{f'}$ contain the residual, i.e. small scales with $\abs{k_x}>\Lambda$. Note that the spatio-temporal mean profile $\boldsymbol{U}$ with $\boldsymbol{k}=(k_x, k_z, \omega) = (0, 0, 0)$ appears as part of $\boldsymbol{\bar{u}}$ under this decomposition. 
%Since QL/GQL includes the full non-linearity in $z$ and time, $\boldsymbol{U}$ offers nothing special with $k_z, \omega = 0$, and behaves the same as all other terms in $\boldsymbol{\bar{u}}$. 

The three terms may be grouped to reflect QL or GQL system formulations. In QL, $\Lambda= 0$, such that $\boldsymbol{\bar{u}}$ represents the base flow and $\boldsymbol{u'}$ the perturbation, while $\boldsymbol{\tilde{u}}$ is zero. For GQL, the base flow consists of all contributions with $k_x \le \Lambda$, i.e. $\boldsymbol{\bar{u}}+\boldsymbol{\tilde{u}}$, and $\boldsymbol{u'}$ is the perturbation.

\begin{figure}
	\centering
	\includegraphics[width=313pt]{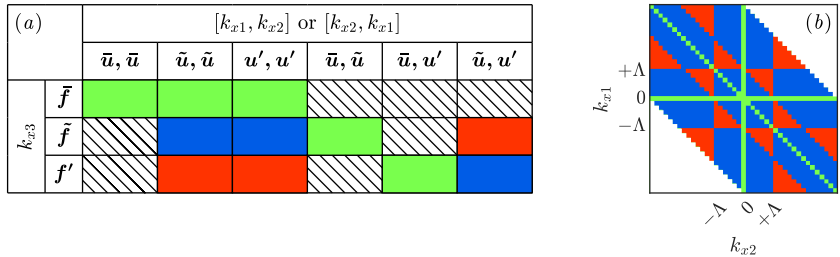}
	\caption{The regions of triadic interactions included in QL/GQL in (\textit{a}) tabular form and (\textit{b}) graphical form for comparison with figure~\ref{fig:pkxom}. The color for the table cells and figure are green for triadic interactions resolved in both QL and GQL; blue for additional triadic interactions included in GQL but not in QL; and red for triadic interactions modeled or neglected in both QL and GQL. Hashed cells in the table indicate prohibited interactions.}
	\label{fig:GQLRegion} 
\end{figure}

Figure~\ref{fig:GQLRegion}(\textit{a}) shows the triadic interactions permitted by QL/GQL in a tabular form, with the six possibilities for the velocity or the velocity gradient listed in the six columns, and the resulting forcing listed in the three rows. These regions of interactions are also plotted in figure~\ref{fig:GQLRegion}(\textit{b}) in a $k_{x1}$ vs $k_{x2}$ plane similar to figure~\ref{fig:pkxom}. In this figure, the color green indicates interactions resolved in both QL and GQL, and corresponds to three lines with $k_{x1}, k_{x2}$, or $k_{x3}=0$ in figure~\ref{fig:GQLRegion}(\textit{b}). The color blue indicates additional interactions included in GQL but not in QL, and in the limiting cases for GQL with $\Lambda = 0$, for which GQL is equivalent to QL, the blue regions disappear, and the triple decomposition collapses to a double decomposition. The color red indicates interactions that must be modeled or neglected in both QL and GQL, and in the limiting case for  GQL with $\Lambda\geq \max(k_x) = 127.5$, the red regions disappear, as all the non-linear interactions included in the DNS are then also included in GQL. Finally, the hashed cells in figure~\ref{fig:GQLRegion}(\textit{a}) indicate non-resonant, prohibited interactions; for example, the interaction of $\boldsymbol{\bar{u}}$ and $\boldsymbol{\bar{u}}$ (both $k_x = 0$) can contribute to $\boldsymbol{\bar{f}}$ ($k_x = 0$), but not $\boldsymbol{\tilde{f}}$ nor $\boldsymbol{f'}$. From figure~\ref{fig:GQLRegion}(\textit{a}), it can be observed that all triadic interactions contributing to $\boldsymbol{\bar{f}}$ are resolved in QL/GQL, while for $\boldsymbol{\tilde{f}}$ and $\boldsymbol{{f'}}$ only part of the triadic interactions are resolved.

Comparing figure~\ref{fig:GQLRegion}(\textit{b}) with figure~\ref{fig:pkxom}(\textit{a}), it can be seen that the QL assumptions do indeed restrict resolved interactions to those corresponding to large interaction coefficients in the DNS. The additional interactions resolved in GQL also correspond to large contributions to the overall forcing. The fractional contribution of GQL-permitted interactions to the total DNS forcing for varying $\Lambda$ can be quantified with the following ratio:
\begin{equation}
	R_{GQL}(\Lambda) = \frac{\sum_{GQL(\Lambda)}  P_{k_x}(k_{x1}, k_{x2})}{\sum_{GQL(\infty)}  P_{k_x}(k_{x1}, k_{x2})}
\end{equation}
where $\sum_{GQL(\Lambda)}$ indicates a summation in the $k_{x1}, k_{x2}$ regions resolved by GQL with the parameter $\Lambda$ (a summation over the green and blue regions in figure~\ref{fig:GQLRegion}). As  $\Lambda \to \infty$, the GQL assumptions admit the equivalent range of interactions to the DNS, with $R_{GQL}(\Lambda \to \infty) = 1$. 

\begin{figure}
	\centering
	\includegraphics[width=339pt]{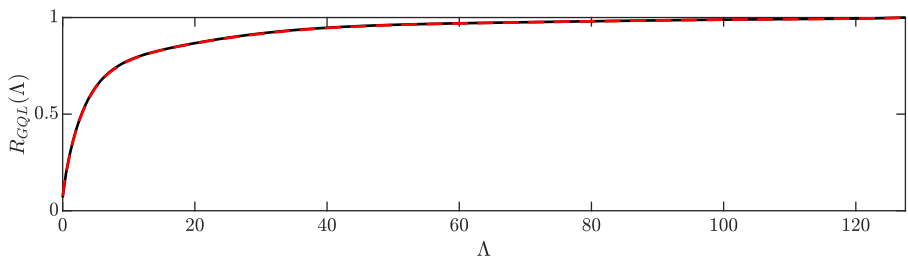}
	\caption{Fraction of total DNS forcing captured by interactions obeying GQL assumptions for various values of $\Lambda$. Temporal averaging performed over 1024 (solid black line) and  32 time snapshots (red dashed line).}
	\label{fig:GQLRatio}
\end{figure}

The ratio $R_{GQL}(\Lambda)$ is plotted in figure~\ref{fig:GQLRatio}, and it can be observed that including a small number of $k_x$ wavenumbers in the base flow using GQL is very effective at capturing the important triadic interactions. 
%In addition, since $P_{k_x}$ is computed through time average, $R_{GQL}$ in figure~\ref{fig:GQLRatio} is computed using different amounts of DNS time snapshots to study its convergence. 
Although the convergence of $P_{k_x}$ requires a significant number of snapshots, $R_{GQL}$ converges rapidly, due to the summation over regions of $P_{k_x}$. Negligible difference is observed in figure~\ref{fig:GQLRatio} when $R_{GQL}$ is computed using the first 32 rather than the full 1024 snapshots, therefore $R_{GQL}$ can be computed with a short statistically steady DNS run.

\section{Summary and Outlook}
In this work, we have characterized spatio-temporal, resonant triadic interactions, which arise due to the quadratic non-linearity in the Navier-Stokes equations viewed from the Fourier domain. We anticipate that this work will be useful in identifying improved modeling of the nonlinearity, especially in quasi-linear, generalized quasi-linear, and resolvent analyses.

We proposed interaction coefficients to quantify the contribution from each pair of interacting wavenumber-frequency combinations to the resulting non-linear forcing. The coefficients show the importance of interactions involving large-scale structures, especially their interaction with the gradients associated with smaller scales.  Interactions involving two smaller scales exciting a larger one or a small scale interacting with the gradients of the larger scale were shown to contribute less, consistent with the physical $y$-support of the smaller scales revealed by spectral analysis.  These types of interaction are consistent with the coherence revealed by amplitude modulation studies which relate large-scale fluctuations to the modulation of near-wall structures. 

Further, the subset of the total interactions that are permitted under QL and GQL reductions correspond well with regions of high amplitude interaction coefficients and increasing the number of GQL-large scales, $\Lambda$, expands the permitted interactions to monotonically increase the total forcing captured. This analysis, performed only on DNS data, reinforces the modeling assumptions underlying QL and GQL approaches, albeit without analyzing the dynamics of QL/GQL. Note that interactions $P_{k_x}(k_{x1}, k_{x2})$ and $P_{k_x}(k_{x2}, k_{x1})$ are not typically differentiated, but have substantially different contributions to $\boldsymbol{f}$.

We emphasize that $R_{GQL}$ is a measurement of the contribution of the interactions permitted under QL/GQL reductions to the total forcing calculated by DNS of the full NSE. As such, it gives an indication of a possible reason for the success of QL and GQL simulations in replicating features of wall turbulence, without consideration of the different dynamics associated with the restricted systems. Further, we have described the importance of triadic interactions by how much they contribute to the total forcing term, $\boldsymbol{f}$. Quantifying their importance in terms of exciting a velocity response would require the use of the linear operator to transform from the forcing domain to the response, for example via equation~\eqref{eq:resolvent_nse}, and/or a Helmholtz decomposition of the forcing to identify the part exciting a velocity response, which are topics of current work.

%\backsection[Supplementary data]{\label{SupMat}Supplementary material and movies are available at \\https://doi.org/10.1017/jfm.2019...}

%\backsection[Acknowledgements]{TBD}

\backsection[Funding]{The support of the U.S. Office of Naval Research under grant numbers N00014-17-1-2960 and N00014-17-1-3022 is gratefully acknowledged. }

\backsection[Declaration of interests]{The authors report no conflict of interest.}

%\backsection[Data availability statement]{The data that support the findings of this study are openly available in [repository name] at http://doi.org/[doi], reference number [reference number]. See JFM's \href{https://www.cambridge.org/core/journals/journal-of-fluid-mechanics/information/journal-policies/research-transparency}{research transparency policy} for more information}

%\backsection[Author ORCIDs]{Y. Huang, https://orcid.org/0000-0002-9457-7964}

%\backsection[Author contributions]{Authors may include details of the contributions made by each author to the manuscript'}

\bibliographystyle{jfm}
\bibliography{./citation}

\begin{thebibliography}{14}
\expandafter\ifx\csname natexlab\endcsname\relax\def\natexlab#1{#1}\fi
\def\au#1{#1} \def\ed#1{#1} \def\yr#1{#1}\def\at#1{#1}\def\jt#1{\textit{#1}}
  \def\bt#1{#1}\def\bvol#1{\textbf{#1}} \def\vol#1{#1} \def\pg#1{#1}
  \def\publ#1{#1}\def\arxiv#1{#1}\def\org#1{#1}\def\st#1{\textit{#1}}

\bibitem[Cheung \& Zaki(2014)]{Cheung14}
{\sc \au{Cheung, L.~C.} \& \au{Zaki, T.}} \yr{2014}  \at{An exact
  representation of the nonlinear triad interaction terms in spectral space}.
  \jt{J. Fluid Mech.}  \bvol{748},  \pg{175--188}.

\bibitem[Duvvuri \& McKeon(2015)]{Duvvuri_McKeon_2015}
{\sc \au{Duvvuri, S.} \& \au{McKeon, B.~J.}} \yr{2015}  \at{Triadic scale
  interactions in a turbulent boundary layer}.  \jt{J. Fluid Mech.}
  \bvol{767}.

\bibitem[Farrell \& Ioannou(2007)]{Farrell_Ioannou_2007}
{\sc \au{Farrell, B.~F.} \& \au{Ioannou, P.~J.}} \yr{2007}  \at{Structure and
  spacing of jets in barotropic turbulence}.  \jt{J. Atmos. Sci.}
  \bvol{64}~(10),  \pg{3652–3665}.

\bibitem[Flores \& Jiménez(2006)]{Flores_Jimenez_2006}
{\sc \au{Flores, O.} \& \au{Jiménez, J.}} \yr{2006}  \at{Effect of
  wall-boundary disturbances on turbulent channel flows}.  \jt{J. Fluid Mech.}
  \bvol{566},  \pg{357–376}.

\bibitem[Gómez {\em et~al.\/}(2014)Gómez, Blackburn, Rudman, McKeon, Luhar,
  Moarref \& Sharma]{Gomez2014}
{\sc \au{Gómez, F.}, \au{Blackburn, H.~M.}, \au{Rudman, M.}, \au{McKeon,
  B.~J.}, \au{Luhar, M.}, \au{Moarref, R.} \& \au{Sharma, A.~S.}} \yr{2014}
  \at{On the origin of frequency sparsity in direct numerical simulations of
  turbulent pipe flow}.  \jt{Phys. Fluids}  \bvol{26}~(10),  \pg{101703}.

\bibitem[Karban {\em et~al.\/}(2023)Karban, Martini, Cavalieri \&
  Jordan]{Karban_NL_23}
{\sc \au{Karban, U.}, \au{Martini, E.}, \au{Cavalieri, A. V.~G.} \& \au{Jordan,
  P.}} \yr{2023}  \at{Modal decomposition of nonlinear interactions in wall
  turbulence}.  \jt{arXiv 2301.01078} .

\bibitem[Lii {\em et~al.\/}(1976)Lii, Rosenblatt \&
  Van~Atta]{Lii_Rosenblatt_Atta_1976}
{\sc \au{Lii, K.~S.}, \au{Rosenblatt, M.} \& \au{Van~Atta, C.}} \yr{1976}
  \at{Bispectral measurements in turbulence}.  \jt{J. Fluid Mech.}
  \bvol{77}~(1),  \pg{45–62}.

\bibitem[Marati {\em et~al.\/}(2004)Marati, Casciola \& Piva]{Marati04}
{\sc \au{Marati, N.}, \au{Casciola, C.~M.} \& \au{Piva, R.}} \yr{2004}
  \at{Energy cascade and spatial fluxes in wall turbulence.}  \jt{J. Fluid
  Mech.}  \bvol{521},  \pg{191--215}.

\bibitem[Marston {\em et~al.\/}(2016)Marston, Chini \&
  Tobias]{Marston_Chini_Tobias_2016}
{\sc \au{Marston, J.~B.}, \au{Chini, G.~P.} \& \au{Tobias, S.~M.}} \yr{2016}
  \at{Generalized quasilinear approximation: Application to zonal jets}.
  \jt{Phys. Rev. Lett.}  \bvol{116}~(21),  \pg{214501}.

\bibitem[Marusic {\em et~al.\/}(2010)Marusic, Mathis \& Hutchins]{marusic2010}
{\sc \au{Marusic, I.}, \au{Mathis, R.} \& \au{Hutchins, N.}} \yr{2010}
  \at{Predictive model for wall-bounded turbulent flow}.  \jt{Science}
  \bvol{329}~(5988),  \pg{193}.

\bibitem[McKeon(2017)]{mckeon2017}
{\sc \au{McKeon, B.~J.}} \yr{2017}  \at{The engine behind (wall) turbulence:
  perspectives on scale interactions}.  \jt{J. Fluid Mech.}  \bvol{817}.

\bibitem[McKeon \& Sharma(2010)]{mckeon2010}
{\sc \au{McKeon, B.~J.} \& \au{Sharma, A.~S.}} \yr{2010}  \at{A critical-layer
  framework for turbulent pipe flow}.  \jt{J. Fluid Mech.}  \bvol{658},
  \pg{336–382}.

\bibitem[Morra {\em et~al.\/}(2021)Morra, Nogueira, Cavalieri \&
  Henningson]{Morra_Nogueira_Cavalieri_Henningson_2021}
{\sc \au{Morra, P.}, \au{Nogueira, P. A.~S.}, \au{Cavalieri, A. V.~G.} \&
  \au{Henningson, D.~S.}} \yr{2021}  \at{The colour of forcing statistics in
  resolvent analyses of turbulent channel flows}.  \jt{J. Fluid Mech.}
  \bvol{907},  \pg{A24}.

\bibitem[Schmidt(2020)]{Schmidt_2020}
{\sc \au{Schmidt, O.~T.}} \yr{2020}  \at{Bispectral mode decomposition of
  nonlinear flows}.  \jt{Nonlinear Dyn.}  \bvol{102}~(4),  \pg{2479–2501}.

\end{thebibliography}
\end{document}